\documentclass[aps,prl,preprint,superscriptaddress]{revtex4-1}
\usepackage{graphicx} 
\usepackage{color} 

\begin{document}

\title{Ice Nucleation on Carbon Surface Supports the Classical Theory
  for Heterogeneous Nucleation}

\author{Raffaela Cabriolu} \affiliation{Department of Civil and
  Environmental Engineering, George Washington University, Washington,
  DC 20052}

\author{Tianshu Li} \email[corresponding
  author:]{tsli@gwu.edu} \affiliation{Department of Civil and Environmental
  Engineering, George Washington University, Washington, DC 20052}



\begin{abstract}

The prevalence of heterogeneous nucleation in nature was explained
qualitatively by the classical theory for heterogeneous nucleation
established over more than 60 years ago, but the quantitative validity
and the key conclusions of the theory have remained
unconfirmed. Employing the forward flux sampling method and the
coarse-grained water model mW, we explicitly computed the
heterogeneous ice nucleation rates in the supercooled water on a
graphitic surface at various temperatures.  The independently
calculated ice nucleation rates were found to fit well according to
the classical theory for heterogeneous nucleation. The fitting
procedure further yields the estimate of the potency factor which
measures the ratio of the heterogeneous nucleation barrier to the
homogeneous nucleation barrier. Remarkably, the estimated potency
factor agrees quantitatively with the volumetric ratio of the critical
nuclei between the heterogeneous and homogeneous nucleation. Our
numerical study thus provides a strong support to the quantitative
power of the theory, and allows understanding ice nucleation behaviors
under the most relevant freezing conditions.

\end{abstract}

\maketitle

The freezing of water nearly all proceeds with the assistance from
foreign substances, a process known as heterogeneous nucleation. In
clouds, the dominant candidates for heterogeneous ice nucleation are
bacteria, pollen grains, mineral dusts, soot particles, and
high-molecular-weight organic compounds
\cite{Cantrell:2005cb,Murray:2012jk}. Despite its ubiquity, the
microscopic picture behind such prevailing process still remains
elusive, because of the complex and stochastic nature of the
heterogeneous nucleation event. In particular, the mechanisms
controlling heterogeneous ice nucleation are not well understood.

Although a molecular understanding is still missing, the thermodynamic
rationale behind the heterogeneous nucleation was already provided in
1950's by the classical theory for heterogeneous nucleation
\cite{Turnbull:1950iq,D:1950fa}, an extension to the classical
nucleation theory (CNT) \cite{Volmer:1926tq} for homogeneous
nucleation, on the basis of macroscopic arguments. According to CNT,
the formation of a critical nucleus needs to overcome a free energy
barrier $\Delta G^*_{\text{hom}}$ through spontaneous fluctuations. In
the case of a spherical solid nucleus forming from the supercooled
liquid, the free energy barrier can be expressed as
\begin{eqnarray}
\Delta G^*_{\text{hom}}=\frac{16\pi \gamma_{ls}^3}{3(\rho \Delta
\mu_{ls})^2} \;, \label{homobarrier}
\end{eqnarray}
where $\gamma_{ls}$ is the solid-liquid interface free energy, $\Delta
\mu_{ls}$ is the chemical potential difference between liquid and
solid, and $\rho$ is the density of liquid. The homogeneous nucleation
rate $R_{\text{hom}}$ varies with the nucleation temperature $T$
following the Arrhenius equation \cite{Kelton}:
\begin{eqnarray}
R_{\text{hom}}=A_{\text{hom}}\;\exp (-\Delta G^{*}_{\text{hom}}/k_BT)
\;, \label{CNT}
\end{eqnarray}  
where $A_{\text{hom}}$ is the kinetic pre-factor. For homogeneous ice
nucleation, both experiments
\cite{Kramer:1999p6333,Koop:2000p611,Stockel:2005p6326,Taborek:1985td,Murray:2010p3808}
and simulations
\cite{Li:2011p7738,Li:2013ko,Moore:2104ku,Reinhardt:2012fz,Sanz:2013hk,Espinosa:2014di}
suggest that the temperature dependence of the homogeneous ice
nucleation rate may be quantitatively described by CNT, with
parametrization being refined by the controlled experiments
\cite{Koop:2009p6666}.

\begin{figure}[t]
\includegraphics[width=2.5 in]{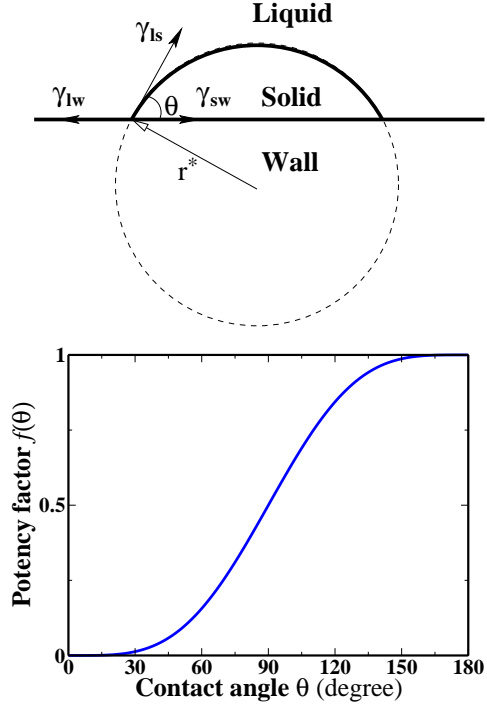}
\caption{ (a) A solid critical nucleus forms at the interface between
  liquid and a foreign flat wall, with a solid contact angle
  $\theta_c$. The contact angle $\theta_c$ can be determined by the
  surface tensions through Eqn. (\ref{Young}). The solid nucleus is
  assumed to be part of a sphere (dashed line) with a radius $r^*$,
  {\em i.e.}, a spherical cap with the volume
  $V_{\text{cap}}=f(\theta_c) V_{\text{sphere}}$. (b) The potency
  factor $f(\theta_c)$ increases from 0 to 1 (homogeneous), as the
  solid contact angle $\theta_c$ varies from 0 to $180^o$. }
\label{Turnbull}
\end{figure}

When a foreign flat wall (W) is present, the solid nucleus can
preferentially form at the interface between the liquid and the wall
(Fig. \ref{Turnbull}(a)). At its critical size, the solid embryo is
under the unstable equilibrium with respect to the dissolution and the
growth, which also indicates a mechanical equilibrium. At the
liquid-solid-wall triple junction, solving the equation of equilibrium
yields the Young's equation
\begin{eqnarray}
\gamma_{lw}=\gamma_{sw}+\gamma_{ls} \cos\theta_c \;, \label{Young}
\end{eqnarray}
where $\gamma_{lw}$ and $\gamma_{sw}$ are the surface tensions for
liquid-wall and solid-wall interfaces. $\theta_c$ defines the contact
angle of solid embryo on the flat wall, with $\theta_c=0$ and
$\theta_c=180^o$ indicating the complete wetting of the wall by solid
and liquid, respectively. If the solid nucleus is further assumed to
be part of the sphere, {\em i.e.}, a spherical cap, its volume
$V_{\text{cap}}$ can be expressed as $V_{\text{cap}}=f(\theta_c)
V_{\text{sphere}}$, where
\begin{eqnarray}
f(\theta_c)=(1-\cos \theta_c)^{2}(2+\cos \theta_c)/4 \;,
\label{potency}
\end{eqnarray} 
and $V_{\text{sphere}}$ is the volume of the sphere containing the
cap. Remarkably, under the framework of classical theory for
heterogeneous nucleation, the factor $f(\theta_c)$ coincides with the
ratio of the free energy barriers between the heterogeneous and
homogeneous nucleation, {\em i.e.}, $f(\theta_c)=\Delta
G^*_{\text{het}}/\Delta G^*_{\text{hom}}$. It thus follows that
\begin{eqnarray}
f(\theta_c)=\frac{\Delta G^*_{\text{het}}}{\Delta
  G^*_{\text{hom}}}=\frac{V_{\text{cap}}}{V_{\text{sphere}}}
  \label{red}
\end{eqnarray}

Eqn. (\ref{red}) provides a simple but robust explanation for the
preference of the heterogeneous nucleation over the homogeneous
nucleation: Instead of forming a spherical nucleus from spontaneous
thermal fluctuations, only part of the sphere $V_{\text{cap}}$ needs
to be nucleated when a foreign surface is present. Accordingly, the
free energy barrier is reduced by the same factor $f(\theta_c)$ by
which the volume of critical nucleus is reduced. Since $f(\theta_c)$
measures the degree of the free energy reduction, it is also known as
the potency factor.  According to Eqn. (\ref{potency}), the potency
factor $f(\theta_c)$ for a foreign wall is determined by the solid
contact angle $\theta_c$, and varies between 0 and 1, as shown in
Fig. \ref{Turnbull}(b). A wall with a lower solid contact angle yields
a lower potency factor $f(\theta_c)$, thus further enhancing
heterogeneous nucleation. Then the heterogeneous nucleation rate can
be expressed by
\begin{eqnarray}
R_{\text{het}}=A_{\text{het}}\;\exp (-f(\theta_c)\Delta
G^{*}_{\text{hom}}/k_BT) \;, \label{het}
\end{eqnarray}
 
Although the classical theory for heterogeneous nucleation offers a
qualitative explanation to the prevalence of heterogeneous nucleation,
its quantitative validity remains unconfirmed.  Auer and Frenkel
\cite{Auer:2003vd} employed umbrella sampling method to calculate the
nucleation barrier of the hard-sphere crystal that completely wets the
smooth walls, and found that the computed barrier height is substantially higher than that predicted by the CNT. The disagreement
was attributed to the omission in the CNT of the line tension at the
liquid-solid-wall triple junction, which may
become non-negligible when crystal completely wets the wall. In a recent study by Winter {\em
  et. al.}\cite{Winter:2009tn}, the total surface energies of the liquid
nuclei (of both spherical and spherical cap shape) were directly
obtained by Monte Carlo simulation for the Ising lattice gas
model. It was found that the obtained surface energies could be
comparable with the capillary approximation employed by CNT, if the
line tension effects at the triple junction are considered.  

In this work, we show that the 
heterogeneous ice nucleation on a graphitic surface indeed supports the
quantitative power of the theory. In particular, the validity of
Eqn. (\ref{red}) and (\ref{het}) is strongly supported from the ice
nucleation rates computed explicitly using the forward flux sampling
method over a wide temperature range. Our work thus provides the first
validation of heterogeneous CNT in the partially wetting regime where the
potency factor is far enough from zero. 

Our molecular dynamics (MD) simulations were carried out using the
coarse-grained model of water (mW) \cite{Molinero:2009p2000}. The
inter-molecular interaction between water and carbon was adopted from
a recent parameterization of the two-body term of the mW model, so
that the strength of the water-carbon interaction reproduces the
experimental contact angle ($86^o$) of water on graphite
\cite{Lupi:2014hh}. The model was recently employed in direct MD
simulations to study the heterogeneous ice nucleation on carbon
surface \cite{Lupi:2014hh,Lupi:2014gj,Reinhardt:2014jt}, where the
nonequilibrium freezing temperature of ice was found to increase due to
the preferential nucleation of ice on carbon surface. Here we employ
the forward flux sampling (FFS) method
\cite{Allen:2006p640,Allen:2006p32} to systematically and explicitly
compute the heterogeneous ice nucleation rates at various temperatures
where spontaneous ice nucleation becomes too slow to occur in direct
simulation. The details of the rate constant calculations can be found
in the Supplementary Materials. Our MD simulation includes 4096 water
molecules and 1008 carbon atoms, in a nearly cubic cell with a
periodic boundary condition. The isobaric-isothermal canonical
ensemble (NPT) with a No$\acute{\mbox{s}}$e-Hoover thermostat was
employed, with a relaxation time of 1 ps and 15 ps for temperature and
pressure, respectively. A time step of 5 fs was used. It should be
noted that while the homogeneous nucleation rate is measured by the
nucleation frequency per unit volume, the heterogeneous nucleation
rate should be characterized by the nucleation frequency per unit
area. However because the simulation volume of liquid is small, and
ice nucleation on carbon surface is strongly preferred, it is
convenient to describe the heterogeneous nucleation rate
$R_{\text{het}}$ on the basis of volume, in order to facilitate a
direct comparison with $R_{\text{hom}}$.

\begin{figure}[t]
\includegraphics[width=3 in]{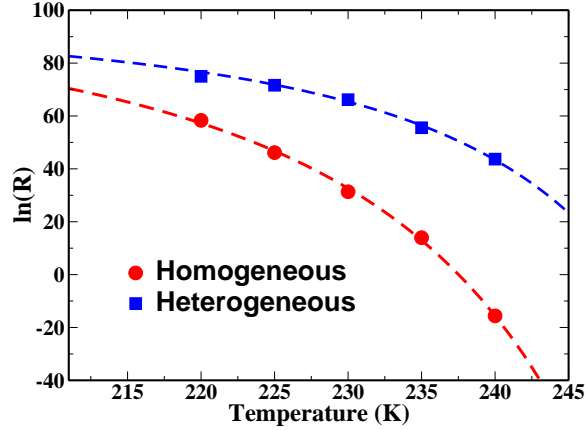}
\caption{Temperature dependence of ice nucleation rate (logarithm) in
  the mW water model, for both homogeneous ice nucleation (red) and
  the heterogeneous nucleation on a graphitic surface (blue). The
  solid red dots and blue squares represent the calculated ice
  nucleation rates by using the FFS method. The dashed red and blue
  lines indicate the fitting, on the basis of the CNT for homogeneous
  nucleation (Eqn. (\ref{lnrate-hom})) and the extension for
  heterogeneous nucleation (Eqn. (\ref{lnrate-het})),
  respectively. The data of homogeneous nucleation were extracted from
  Ref. \cite{Li:2011p7738}. }
\label{rates}
\end{figure}

Figure \ref{rates} shows the computed heterogeneous ice nucleation
rates (in logarithm) as a function of nucleation temperature, in the
range of 220~K to 240~K.  To quantitatively explore the catalytic
activity of the graphitic surface, we compare the obtained
heterogeneous ice nucleation rates with the reported homogeneous ice
nucleation rates from the previous work \cite{Li:2011p7738} using the
FFS method and the mW water model, as shown in Figure \ref{rates}. It
is clear that a graphitic surface yields the significantly enhanced
ice nucleation rates, under all temperatures studied. Our results thus
support the finding by Lupi {\em at. al.} \cite{Lupi:2014hh} and
confirm the enhanced ice nucleation capacity of carbon surface.

The calculated heterogeneous ice nucleation rates at various
temperatures allow assessing the quantitative validity of
Eqn. (\ref{het}). To do this, we fit the obtained heterogeneous ice
nucleation rates at various temperatures according to the theory of
nucleation, using the procedure employed by Li {\em at. al.} in
analyzing the homogeneous ice nucleation \cite{Li:2011p7738}. In this
procedure, the chemical potential difference $\Delta \mu$ is
approximated as a linear function of temperature, {\em i.e.}, $\Delta
\mu=H (T-T_m)/T_m$, where $T_m$ is the equilibrium melting temperature
(274.6~K) of ice in the mW model, and $H$ is a constant; The
liquid-solid interface energy $\gamma_{ls}$ is assumed to be
temperature independent. It is noted that both assumptions have been
verified for the mW water model in different simulation studies
\cite{Espinosa:2014di,Jacobson:2009p6277}. For homogeneous ice
nucleation, it was shown that the independently calculated homogeneous
ice nucleation rate can be fitted according to the following
expression:
\begin{eqnarray}
\ln
(R_{\text{hom}})=\ln(A_{\text{hom}})+\frac{C_{\text{hom}}}{(T-T_m)^2T}
\;,
\label{lnrate-hom}
\end{eqnarray}   
where $\ln (A_{\text{hom}})=114.07\pm 1.86$ and $C_{\text{hom}}=-16\pi
\gamma_{ls}^3 T_m^2/(3 k_B \rho^2 H^2)=-3.72\pm 0.08\times 10^7$ K$^3$
are the fitting constants \cite{Li:2011p7738}. We note that the
nucleation barrier $\Delta G^*_{\text{hom}}=k_B
C_{\text{hom}}/(T-T_m)^2$. The fitting yielded an estimate of
$\gamma_{ls}=31.01\pm 0.21$ mJ m$^{-2}$, which agrees well with the
surface tensions computed through other approaches for the mW water
model \cite{Limmer:2012gw,Espinosa:2014di}.

Remarkably, the obtained heterogeneous ice nucleation rates are also
found to fit well the classical theory for heterogeneous nucleation,
as shown in Figure \ref{rates}. Specifically, the calculated ice
nucleation rates $R_{\text{het}}$ can be fitted according to
\begin{eqnarray}
\ln
(R_{\text{het}})=\ln(A_{\text{het}})+\frac{C_{\text{het}}}{(T-T_m)^2T}
\;.
\label{lnrate-het}
\end{eqnarray}
The fitting yields the estimate of the kinetic pre-factor for
heterogeneous nucleation $\ln (A_{\text{het}})=102\pm 7.70$, which
is consistent with that for the homogeneous nucleation $\ln
(A_{\text{hom}})=114.07\pm 1.86$ \cite{Li:2011p7738}.  More
importantly, the other fitting constant $C_{\text{het}}=-1.70\pm 0.07
\times 10^7$ K$^3$ allows estimating the reduction of the nucleation
barrier, as $C_{\text{het}}/C_{\text{hom}}=\Delta
G^*_{\text{het}}/\Delta G^*_{\text{hom}} \equiv f_b(\theta_c)$. By
comparing the fitting constants $C$ from the heterogeneous and
homogeneous ice nucleation, we obtain the potency factor for the
graphitic surface $f_b(\theta_c)=0.456\pm 0.019$. This corresponds to
a solid contact angle of $\theta_c \sim 86.6^o$. It is noted here that
$\theta_c$ is the contact angle between ice and graphene, and should
not be confused with the water-graphene contact angle $86^o$, although
their magnitudes coincide here.  

\begin{figure}[!]
\includegraphics[width=3.2 in]{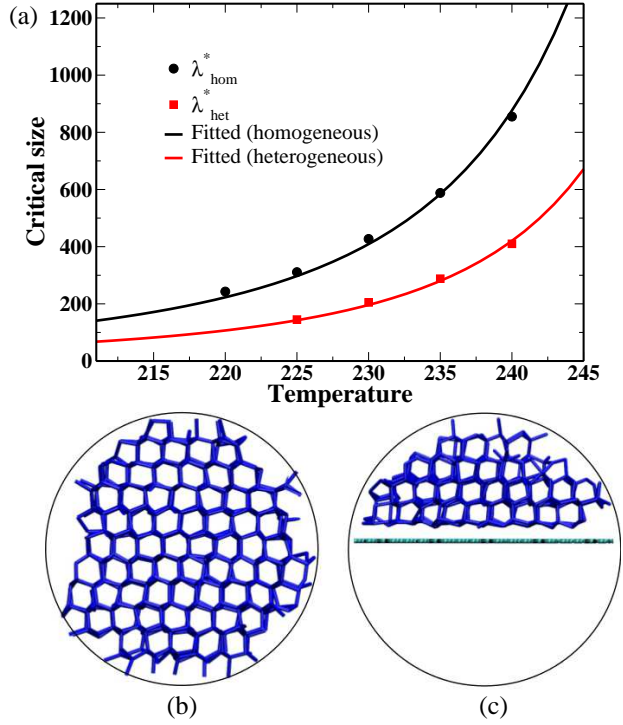}
\caption{(a) Variation of the critical ice nucleus size with temperature
  for both homogeneous and heterogeneous nucleation. The critical size
  is estimated based on the committor probability analysis, which
  yields $\lambda^*_{\text{het}}$: $145\pm 5$ at 225~K, $205\pm 10$ at
  230~K, $288\pm 7$ at 235~K, and $410\pm 10$ at 240~K, respectively
  (See Supplementary Materials for more details). The critical size for
  homogeneous nucleation was obtained in Ref. \cite{Li:2011p7738}. The
  simulation data and the fitted curves are represented by data points
  and solid lines, respectively. (b) A snap shot of the critical ice
  nucleus forming from homogeneous ice nucleation at 240~K. (c) A snap
  shot of the critical ice nucleus forming on graphene surface at 240~K.}
\label{criticalsize}
\end{figure}

It is then of interest to further test the validity of
Eqn. (\ref{red}), namely, the potency factor $f_b(\theta_c)$ can be
also quantitatively related to the volumetric ratio $f_v(\theta_c)$ of
the critical nucleus of the heterogeneous nucleation and the
homogeneous nucleation. We note that such verification becomes
possible in our study because the size of the critical nucleus
$\lambda^*$ can be independently estimated from the ensemble of
nucleation trajectories obtained in the FFS calculation. Using the
definition that the critical nucleus has the equal probabilities of
dissolving and growing completely, {\em i.e.}, with a committor
probability $p_{\mbox{B}}=0.5$ \cite{Bolhuis:2002p54}, we obtained the
estimate of the critical nucleus size $\lambda$ (number of water
molecules contained in the critical ice nucleus) at various nucleation
temperatures, as shown in Fig. \ref{criticalsize}. According to CNT,
the critical size of the spherical nucleus in homogeneous nucleation
is expressed by $\lambda^*_{\text{hom}} =32\pi \gamma^3/(3\rho^2
(\Delta \mu)^3)$. For mW water model, since $\gamma$ is nearly
temperature independent and $\Delta \mu=H (T-T_m)/T_m$
\cite{Espinosa:2014di}, the critical size $\lambda^*_{\text{hom}}$
exhibits the following temperature dependence:
\begin{eqnarray}
\lambda^*_{\text{hom}}(T)=\frac{32\pi \gamma^3}{3\rho^2
  H^3}\frac{1}{\left(\frac{T_m}{T}-1
  \right)^3}=\frac{B_{\text{hom}}}{\left(\frac{T_m}{T}-1 \right)^3}\;,
\label{lambda-hom} 
\end{eqnarray}
where $B_{\text{hom}}$ is the temperature independent constant. The
obtained critical size $\lambda^*_{\text{hom}}$ at various
temperatures are found to fit well according to Eqn
(\ref{lambda-hom}), as shown in Fig. \ref{criticalsize}(a). The good fit
is not unexpected because previous studies
\cite{Reinhardt:2012fz,Li:2013ko} have shown the critical ice nucleus
from homogeneous nucleation is nearly spherical.  The fitting
procedure yields the constant $B_{\text{hom}}=1.752\pm 0.027$. For
the heterogeneous ice nucleation, the same fitting procedure was found
to equivalently apply to the calculated critical nucleus size
$\lambda^*_{\text{het}}$, through
\begin{eqnarray}
\lambda^*_{\text{het}}(T)=\frac{B_{\text{het}}}{\left(\frac{T_m}{T}-1
\right)^3}\;,
\label{lambda-het} 
\end{eqnarray}
which yields the constant $B_{\text{het}}=0.840\pm0.014$. By comparing
the two fitting constants $B$, one obtains the volumetric ratio
$f_v(\theta_c)\equiv\lambda^*_{\text{het}}
/\lambda^*_{\text{hom}}=B_{\text{het}}/B_{\text{hom}}$.  Remarkably,
the obtained volumetric ratio $f_v(\theta_c)=0.480\pm 0.011$ agrees
{\em quantitatively} with the potency factor $f_b(\theta_c)=0.456\pm
0.019$ estimated from the nucleation barriers. The quantitative
validity of Eqn. (\ref{red}), an important conclusion from CNT and its
extension, is thus strongly supported through the molecular simulation
results based on the mW water model.

\begin{figure}[!]
\includegraphics[width=3 in]{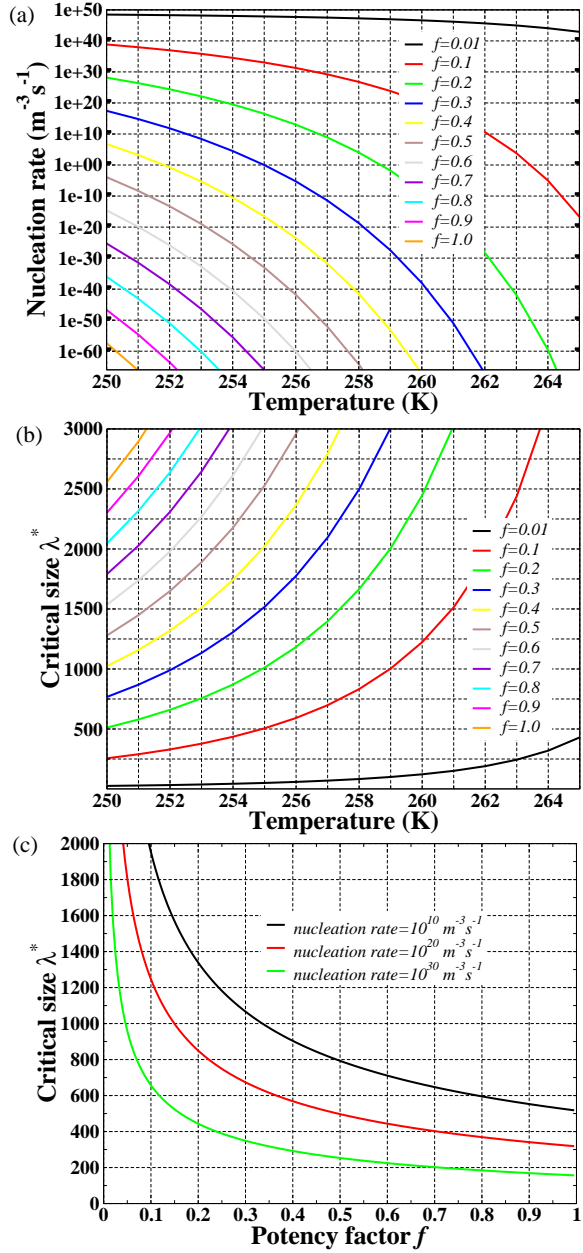}
\caption{The predicted temperature variation of (a) ice nucleation
rate $R(T)$ and (b) critical size of ice nucleus $\lambda^*(T)$ in the
mW water model, due to the presence of a heterogeneous nucleation
center with different potency factor $f$. A small potency factor $f$
indicates a strong ice nucleation efficacy, and $f=1$ corresponds to
homogeneous ice nucleation. (c) The predicted variation of the
critical size $\lambda^*$ for different potency factor $f$, subject to
a fixed nucleation rate. }
\label{prediction}
\end{figure}

The verified quantitative validity of the CNT (and its extension) then
allows predicting ice nucleation behavior in the presence of a
heterogeneous nucleation center. The nucleation efficacy of the
foreign surface can be generally described based on its potency factor
$f$. Using the fitted kinetic pre-factor $A_{\text{hom}}$ ($\approx
A_{\text{het}}$) and $C_{\text{hom}}$, and Eqn. (\ref{lnrate-het}),
one obtains both the heterogeneous ice nucleation rate and the
corresponding critical nucleus size, as a function of the nucleation
temperature, for different potency factors $f$. As shown in
Fig. \ref{prediction}, the predicted ice nucleation rates clearly
indicate the preference and the relevance of the heterogeneous
nucleation at the moderate and low supercooling. For example, at
250~K, a nucleation center with the potency factor of $f=0.1$
(equivalent to a solid contact angle $\theta_c=52.5^o$) yields an ice
nucleation rate about 100 orders of magnitude higher than that of
homogeneous nucleation at the same temperature. Intriguingly, the
sizes of the critical nuclei from these relevant nucleation events
fall within the range of a few hundred to a few thousand water
molecules. This implies that the most relevant ice nucleation events
mediated by an effective nucleation center under a low supercooling,
{\em i.e.}, $-10^o \text{C} \sim -20^o \text{C}$, can be possibly
modeled by molecular simulations through using a reasonable number
($10^3\sim 10^4$) of water molecules.

Interestingly, when {\em ice nucleation rate $R$ is fixed}, the theory of
nucleation predicts that the critical size $\lambda^*$ decreases with
the potency factor $f$ (Fig. \ref{prediction}(c)), and a homogeneous
nucleation yields the minimum critical nucleus. This may appear
surprising, but the prediction can be understood by the fact that the
heterogeneous nucleation producing the same ice nucleation rate occurs
at a much elevated temperature (Fig. \ref{prediction}(a)). For
example, an ice nucleation rate of
$10^{20}\;\mbox{m}^{-3}\mbox{s}^{-1}$ would require a homogeneous
nucleation temperature $T_{\text{hom}} \approx 225$~K, but a
heterogeneous nucleation temperature $T_{\text{het}} \approx 260$~K
for the nucleation center with a potency factor $f=0.1$. The
prediction (Fig. \ref{prediction}(c)) shows the critical nucleus for
such heterogeneous nucleation at 260~K contains about 1220 water
molecules ({\em i.e.}, 1/10 of the critical size of the homogeneous
nucleation at 260~K), larger than the critical size (310) of the
homogeneous nucleation at 225~K.  As the solid contact angle
$\theta_c$ decreases with the nucleation efficacy
(Fig. \ref{Turnbull}), a strong ice nucleation center yields a more
``flat'' ice nucleus that appears increasingly two-dimension
like. It is noted that in such scenario the possible effect from
the line tension at the triple junction can be non-negligible
\cite{Auer:2003vd,Winter:2009tn}. However its quantitative effect in
ice nucleation rate is unclear. Using the density of ice, one can estimate the radius of the
spherical segment ({\em i.e.}, the frustum of the spherical cap) to be
of the order of a few nano meters, implying that the dimension of an
effective ice nucleation site is typically of the order of $10^1$
nm. This estimate may be used as an important parameter in experiments
for potentially observing ice nucleation {\em in situ} and designing
effective strategy for controlling ice nucleation.

\begin{acknowledgments}
The authors thank V. Molinero for valuable discussion. The authors
acknowledge support from the ACS Petroleum Research Fund,
 NSF (Grant No. CBET-1264438), and Sloan Foundation through the Deep
 Carbon Observatory. 
\end{acknowledgments}



%

\end{document}


\title{Supplementary Materials for ``Ice Nucleation on Carbon Surface Supports the Classical Theory for Heterogeneous Nucleation''}

\author{Raffaela Cabriolu} \affiliation{Department of Civil and
  Environmental Engineering, George Washington University, Washington,
  DC 20052}

\author{Tianshu Li} \email[corresponding
  author:]{tsli@gwu.edu} \affiliation{Department of Civil and Environmental
  Engineering, George Washington University, Washington, DC 20052}

\maketitle

\section{Calculation of ice nucleation rates}

\begin{figure}[h]
\includegraphics[width=5 in]{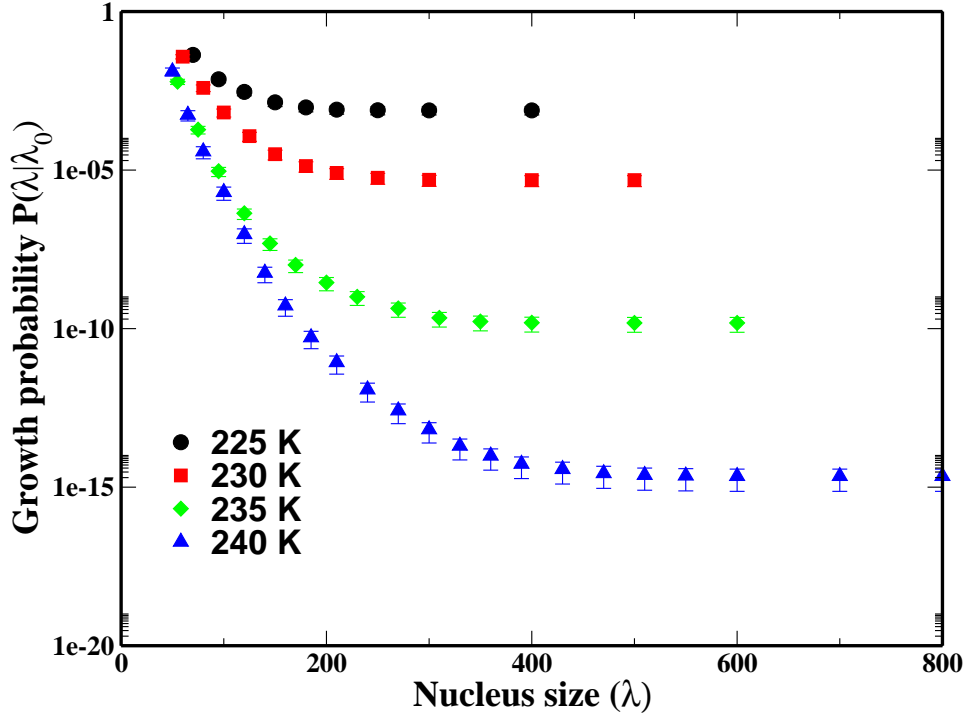}
\caption{Calculated growth probability $P(\lambda |\lambda_0)$ as a
  function of ice nucleus size $\lambda$ for heterogeneous ice
  nucleation on a graphitic surface at 225~K, 230~K, 235~K, and
  240~K. The interfaces $\lambda_i$ used in FFS simulation correspond
  to the abscissa of the data points. }
\label{growth}
\end{figure}

We employed the forward flux sampling (FFS) method \cite{Allen:2006p32} to compute the ice
nucleation rate on graphitic surface. The method has been applied to
successfully study the
homogeneous ice nucleation \cite{Li:2011p7738} using the mW water model. In FFS,
the rate constant for the transition from the basin $A$ to the basin $B$ is
obtained through the ``effective positive flux'' expression \cite{vanErp:2003p1941}:
$R_{AB}=\dot{\Phi}_{\lambda_0} P(\lambda_B|\lambda_0)$, where
$\dot{\Phi}_{\lambda_0}$ and $P(\lambda_B|\lambda_0)$ are the initial flux
rate crossing the first interface $\lambda_0$ from basin $A$, and the
probability for a trajectory starting from $\lambda_0$ to reach basin
$B$, respectively. The interfaces are defined based on an order parameter $\lambda$,
{\em i.e.}, number of water molecules contained in the largest ice
cluster. The ice-like and liquid-like water molecules are identified
based on the local order parameter $q_6$ \cite{Li:2011p7738}, and an
ice-like water molecule is defined as the one with $q_6>0.5$. More
details of computing $q_6$ can be found in
Ref. \cite{Li:2011p7738}. The initial flux rate can be computed by
direct molecular dynamics simulation, through
$\dot{\Phi}_{\lambda_0}=N_0/(t_0V)$, where $N_0$ is the
number of successful crossings to the interface $\lambda_0$ from basin
$A$, $t_0$ is the total time of the initial sampling, and $V$ is the
simulation volume. As our recent work \cite{Bi:2014bt} has demonstrated
the importance of the initial sampling on the reliability of the final
nucleation rate, particularly in a heterogeneous environment, a
sufficiently large sampling time ($t_0>300$ ns) is used for computing
$\dot{\Phi}_{\lambda_0}$ to ensure the convergence. The growth
probability $P(\lambda_B|\lambda_0)$ is obtained through
$P(\lambda_B|\lambda_0)=\prod_{i=1}^{n}P(\lambda_i|\lambda_{i-1})$ and
the individual crossing probability
$P(\lambda_i|\lambda_{i-1})=N_i/M_{i-1}$, where $N_i$ is the number of
successful crossings ($\sim 110$) to the interface $\lambda_i$, and $M_{i-1}$
is the number of trial shootings (typically of $10^3$) from the
interface $\lambda_{i-1}$.  The more details for computing
$P(\lambda_B|\lambda_0)$ are explained in
Ref. \cite{Li:2009p4363,Li:2011p7738}. The statistical uncertainty of
nucleation rate is mainly attributed to the error in the calculated
$P(\lambda_i|\lambda_{i-1})$, which includes the variance of
the binomial distribution of $N_i$ and the landscape variance of the
configurations collected at the previous interface $\lambda_{i-1}$
\cite{Allen:2006p32}. The calculated initial flux rate
$\dot{\Phi}_{\lambda_0}$, growth probability $P(\lambda_B|\lambda_0)$,
and the final nucleation rates $R$ are listed in Table \ref{ratesall}
for the heterogeneous ice nucleation at 225~K, 230~K, 235~K, and 240~K.
The computed growth probability $P(\lambda |\lambda_0)$, as a function
of nucleus size $\lambda$, is also shown in Fig. \ref{growth}.     

\begin{table}[!]
\caption{Calculated initial flux rate $\dot{\Phi}_{\lambda_0}$, growth
probability $P(\lambda_B|\lambda_0)$, and nucleation rate $R$ for
heterogeneous ice nucleation on a graphitic surface, using the mW water
model and the FFS method. The critical nucleus size $\lambda^*$ is
determined by the committor probability analysis, on the basis of the
computed growth probability.}
\begin{tabular}{ccccc}
T/K & $\dot{\Phi}_{\lambda_0}$/$\mbox{m}^{-3}\mbox{s}^{-1}$
& $P(\lambda_B|\lambda_0)$ & Nucleation rate $R$
/$\mbox{m}^{-3}\mbox{s}^{-1}$ & Critical size $\lambda^*$ \\ \hline
225 & $1.69 \times 10^{34}$ & $7.64\pm 2.16 \times 10^{-4}$ & $1.29\pm
0.36 \times 10^{31} $ & $145\pm 5$ \\
230 & $1.12 \times 10^{34}$ & $4.83 \pm 1.83 \times 10^{-6}$ & $5.41\pm
2.05 \times 10^{28} $ & $205\pm 10$ \\
235 & $8.90 \times 10^{33}$ & $1.51\pm 0.74 \times 10^{-10}$ & $1.34\pm
0.66 \times 10^{24} $ & $288\pm 7$ \\
240 & $4.19 \times 10^{33}$ & $2.23\pm 1.49 \times 10^{-15}$ & $9.34\pm
6.24 \times 10^{18} $ & $410\pm 10$ \\
\label{ratesall}
\end{tabular}
\end{table}

At 220~K, ice nucleation becomes accessible in direct molecular
dynamics simulation. Three out of four trajectories lead to
spontaneous ice crystallization, with a total simulation time around
30 ns, which yields an estimate of the nucleation rate of the order of
$10^{32}\; \mbox{m}^{-3}\mbox{s}^{-1}$. 

\begin{figure}[!]
\includegraphics[width=5 in]{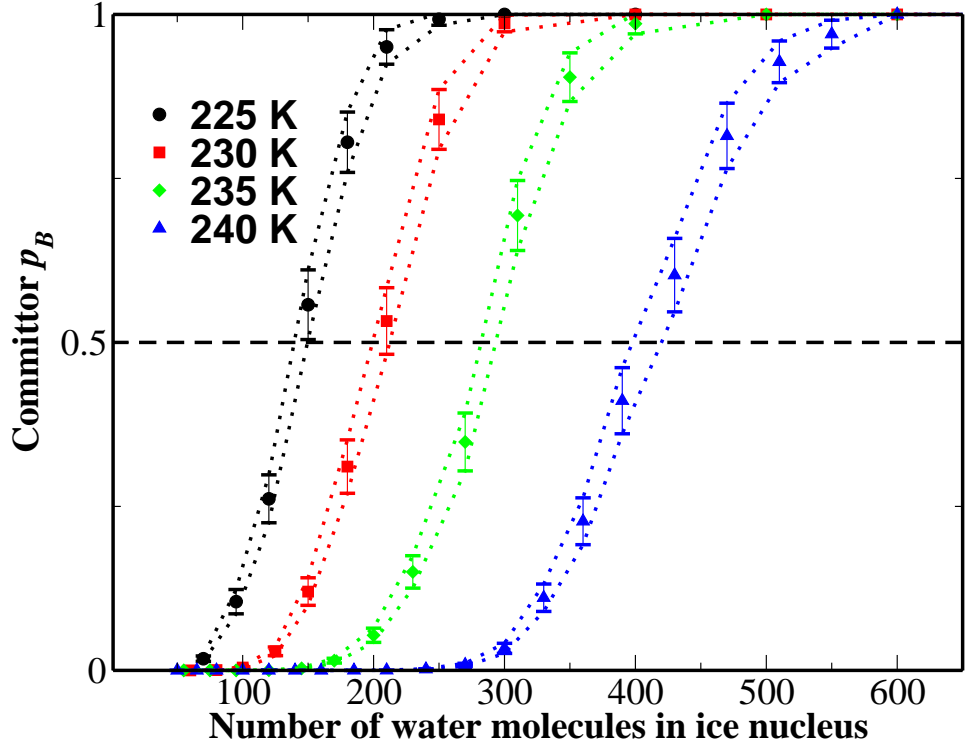}
\caption{Calculated committor $p_{\mbox{B}}$ as a function of the
  nucleus size $\lambda$ for heterogeneous ice nucleation on a
  graphitic surface at various nucleation temperatures. The dotted
  lines indicate the upper and lower boundaries of the calculated
  $p_{\mbox{B}}$, on the basis of the estimated error bar of
  $p_{\mbox{B}}$. The horizontal dash line corresponds to a
  $p_{\mbox{B}}=0.5$, and intersects the both boundaries of the
  calculated $p_{\mbox{B}}$, which yields the estimate of the
  uncertainty of $\lambda^*$.  }
\label{committor}
\end{figure}

The size of the critical ice nucleus $\lambda^*$ can be numerically determined by
the committor probability $p_{\mbox{B}}$ \cite{Bolhuis:2002p54}. A critical crystalline nucleus is
defined as the one with 50\% of probability of growing completely into
crystal, {\em i.e.}, $p_{\mbox{B}}=0.5$. The committor $p_{\mbox{B}}$ as a function
of nucleus size $\lambda$ is obtained on the basis of the computed
growth probability $P(\lambda_i|\lambda_{i-1})$, and the critical size
$\lambda^*$ under different temperatures are determined
(Fig. \ref{committor}) and listed in Table \ref{ratesall}.

\begin{figure}[!]
\includegraphics[width=5 in]{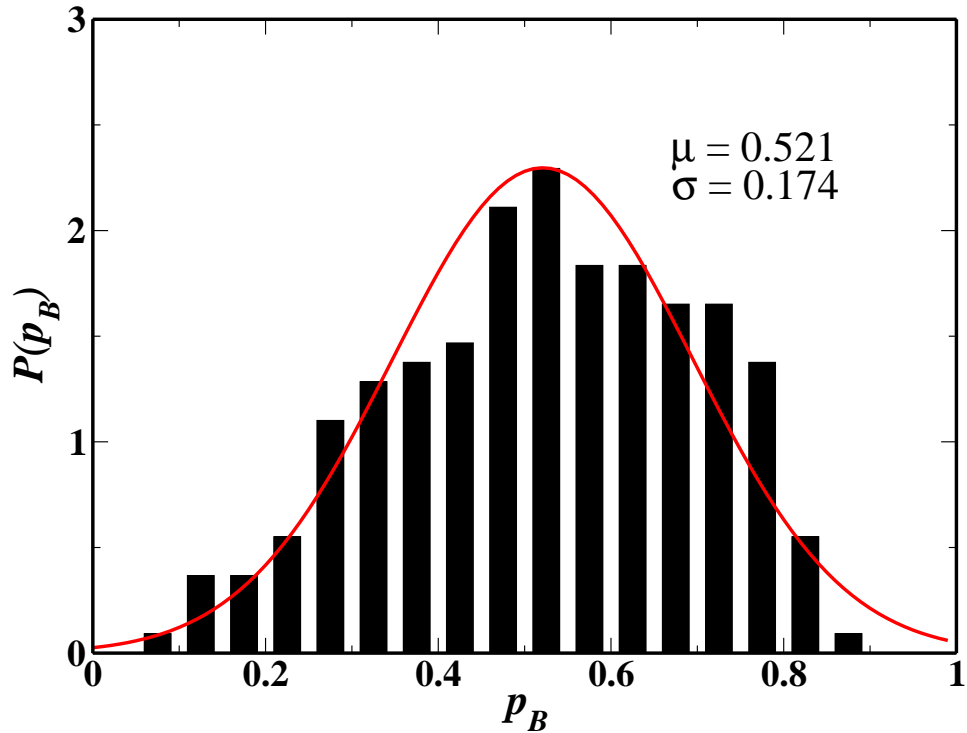}
\caption{The computed histogram of committor $p_{\mbox{B}}$ for
  heterogeneous ice nucleation at 230~K at the interface
  $\lambda=210$. The Gaussian distribution function with the same
  intrinsic mean and standard deviation is shown as the red line.}
\label{committor-dist}
\end{figure}

To understand whether the size of the nucleus $\lambda$ is a good
approximation to the actual reaction coordinate, we carried out the committor
distribution analysis near the critical size $\lambda^*\approx 205$
for the heterogeneous nucleation at 230~K. 110 configurations were
selected from the interface $\lambda=210$ and each one receives 20
shootings with Boltzmann distributed velocities. The obtained
committor distribution is shown in Fig. \ref{committor-dist}, along
with the Gaussian distribution with the intrinsic mean $\mu$ and standard
deviation $\sigma$ \cite{Peters:2006cv}. The calculated distribution
is peaked at $\mu=0.521$, with an overall agreement with the intrinsic
committor distribution, indicating that $\lambda$ fairly accurately
describes the actual reaction coordinate, and is thus a good order
parameter.   

%